\documentclass[superscriptaddress,twocolumn,english,floatfix,aps,prl,amsmath,amssymb,longbibliography]{revtex4-2}
\usepackage[T1]{fontenc}
\usepackage[latin9]{inputenc}
\usepackage{babel}
\usepackage{polski}
\usepackage{comment}
\usepackage{color}
\usepackage{times}
\usepackage{epsfig}
\usepackage{subfigure}
\usepackage{amsmath}
\usepackage{amssymb}
\usepackage{graphicx}
\usepackage[colorlinks=true]{hyperref}
\hypersetup{citecolor=blue,linkcolor=blue,urlcolor=blue}

\begin{document}

\title{Peak, valley and intermediate regimes in the lateral van der Waals force}

\author{Edson C. M. Nogueira}
\email{edson.moraes.nogueira@icen.ufpa.br}
\affiliation{Faculdade de F\'{i}sica, Universidade Federal do Par\'{a}, 66075-110, Bel\'{e}m, Par\'{a}, Brazil}

\author{Lucas Queiroz}
\email{lucas.silva@icen.ufpa.br}
\affiliation{Faculdade de F\'{i}sica, Universidade Federal do Par\'{a}, 66075-110, Bel\'{e}m, Par\'{a}, Brazil}

\author{Danilo T. Alves}
\email{danilo@ufpa.br}
\affiliation{Faculdade de F\'{i}sica, Universidade Federal do Par\'{a}, 66075-110, Bel\'{e}m, Par\'{a}, Brazil}

\date{\today}

%%%%%%%%%%%%%%%%%%%%%%%%%%%%%%%%%%%%%%%%%%%%%%%%%%%%%%%%%%%%%%%
\begin{abstract}
We study the van der Waals (vdW) interaction between a polarizable particle and a grounded conducting corrugated surface.
For sinusoidal corrugations, one knows that, under the action of the lateral vdW force, an isotropic particle is always attracted to the nearest corrugation peak, with such behavior called in the present paper as peak regime.
Here, considering an anisotropic polarizable particle, and making analytical calculations valid beyond the proximity force approximation (PFA), we show that the attraction is not only toward the peaks, but, for certain particle orientations and distances from the surface, the lateral force attracts the particle to the nearest corrugation valley (valley regime), or even to an intermediate point between a peak and a valley (intermediate regime).
We also show that in the configurations of transition between the peak and valley regimes the lateral vdW force vanishes, even in the
presence of a corrugated surface.
In addition, we find that these new regimes occur in general, for periodic and nonperiodic corrugated surfaces.
Moreover, we demonstrate that similar regimes arise in the classical interaction between a neutral polarized particle and a rough surface.
The description of these valley and intermediate regimes, which are out of reach of the predictions based on the PFA, 
may be relevant for a better understanding of the interaction between anisotropic particles and corrugated surfaces in classical and quantum physics, with experimental verifications feasible in both domains.  
\end{abstract}

\maketitle

%%%%%%%%%%%%%%%%%%%%%%%%%%%%%%%%%%%%%%%%%%%%%%%%%%%%%%%%%%%%%%%%%%%%%%%%%%%%%%%%%%%%%%%%%%%
\section{Introduction}
\label{sec-intro}

In the 1940s decade, Casimir and Polder
calculated the interaction between a neutral atom and a perfectly conducting plate, taking into account the retardation of the electromagnetic interaction \cite{Casimir-PhysRev-1948}.
In the limit of very small distances between the atom and the plate, if compared to the wavelengths of the atomic transitions, the Casimir-Polder (CP) formula reduces to the London-van der Waals (vdW) interaction between the atom and the plate \cite{Casimir-PhysRev-1948}.
In general, CP/vdW dispersion forces depend on the geometry of the material bodies and on the boundary conditions that they impose on the electromagnetic field \cite{Israelachvili-PRSL-1972,Milonni-QuantumVacuum-1994,Bordag-Book-2009,Israelachvili-Book-2011,Buhmann-DispersionForces-I,Buhmann-DispersionForces-II,Passante-Symmetry-2018}. These forces have important consequences in physics, chemistry,  biology and engineering \cite{Dimopoulos-PRD-2003,Parsegian-Book-2006,Woods-RMP-2016},
and the progress in the precision of the experiments that measure them has opened possibilities for many applications in micro and nanotechnology \cite{Ball-Nature-2007,Rodriguez-Capasso-PRL-2010,Rodriguez-NaturePhotonics-2011,Keil-JourModOpt-2016}.

In the context of the CP/vdW interaction, the interest in the subtleties brought by the consideration of anisotropic polarizable particles has increased \cite{Shajesh-PRA-2012,Buhmann-PRA-2018,Bimonte-PRD-2015,Thiyam-PRA-2015,Gangaraj-PRB-2018,Antezza-PRB-2020,Buhmann-IJMPA-2016, Venkataram-PRA-2020,Eberlein-PRA-2011}.
Among the reasons is the phenomenon known as the Casimir torque 
\cite{Bimonte-PRD-2015,Thiyam-PRA-2015,Gangaraj-PRB-2018,Antezza-PRB-2020},
and the prediction of a vertical repulsive CP/vdW force involving 
anisotropic particles \cite{Levin-PRL-2010,Buhmann-IJMPA-2016,Venkataram-PRA-2020,Eberlein-PRA-2011}.
%Eberlein-PRA-2011
To take into account curvature or roughness effects in the interaction with these particles, theoretical analyses beyond the proximity force approximation (PFA) are necessary when such effects  depend on a more precise description of the surface geometry \cite{Bimonte-PRD-2015}.
During the last two decades, several studies have been done using approaches that can provide predictions for physical situations for which the PFA is not the best approximation
\cite{Gies-JHEP-2003, Neto-PRA-2005, Emig-PRL-2006, Bimonte-EPL-2012,Bimonte-APL-2012,Lussange-PRA-2012,Dalvit-PRL-2008,Bimonte-PRD-2015,Bennett-PRA-2015, Buhmann-IJMPA-2016}.
For instance, for isotropic particles interacting with
grooved surfaces, the PFA predicts that no lateral force acts on the particle when it is over a plateau of the surface, whereas analyses beyond the PFA reveals the
existence of a lateral force \cite{Dalvit-PRL-2008}.

In the present paper, combining the perturbative analytical solution for the Green function obtained by Clinton, Esrick, and Sacks \cite{Clinton-PRB-1985} (related to Poisson's equation of a point charge in the presence of an ideal conducting nonplanar surface) with the description done by Eberlein and Zietal \cite{Eberlein-PRA-2007} (for the vdW interaction between a polarizable particle and an ideal conducting surface), we propose 
a new analytical approach to investigate the vdW interaction between an anisotropic particle and an ideal grounded conducting corrugated surface.
This approach requires no demand on the smoothness of the rough surface, so that our results are valid beyond the PFA.  
In the sequence, we apply our formulas to a sinusoidal surface with corrugation period $\lambda$, distant $z_0$ from the particle, and predict that, under the action of the lateral vdW force, a neutral isotropic particle is always attracted to the nearest corrugation peak, in agreement with results found in the literature \cite{Bezerra-PRA-2000,Dalvit-PRL-2008}, a behavior called in the present paper as peak regime.
On the other hand, for the case of a neutral anisotropic polarizable particle, 
we show that, depending on the ratio $\lambda/z_0$ and particle orientation, it can also be attracted to a corrugation valley (valley regime), or to a point between a peak and a valley (intermediate regime).
In all situations, taking the limit $\lambda/z_0\rightarrow\infty$, our formulas lead to
results according to the PFA, for which we show that, even considering anisotropic particles, only the peak regime is predicted.
We also show that similar regimes (valley and intermediate) also arise in the classical interaction between a neutral polarized particle and a sinusoidal surface, thus providing additional experimental possibilities to verify the existence of these new regimes.

%%%%%%%%%%%%%%%%%%%%%%%%%%%%%%%%%%%%%%%%%%%%%%%%%%%%%%%%%%%%%%%%%%%%%%%%%%%%%%%%%%%%%%%%%%
\section{Our Approach}
\label{sec-used-approach}

\begin{figure}[h]
	\centering
	\epsfig{file=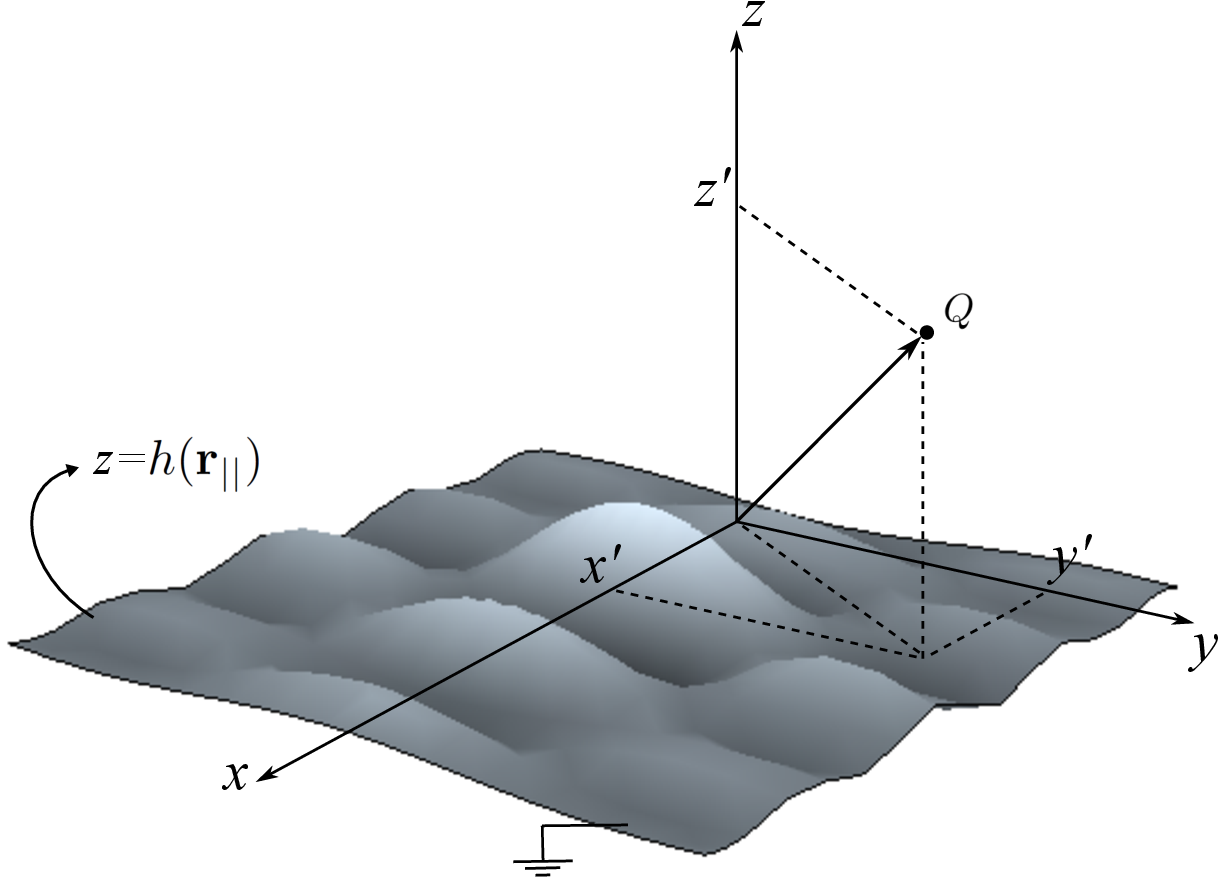,  width=0.8 \linewidth}
	\caption{
		Illustration of a charge $Q$, located at ${\bf r}^{\prime}=x^\prime\hat{{\bf x}}+y^\prime\hat{{\bf y}}+z^\prime\hat{{\bf z}}$
		(with $z^\prime>0$), interacting with a general grounded conducting corrugated surface, whose corrugation profile is described by $z=h(\textbf{r}_\parallel)$.
	}
	\label{fig:carga-superficie-geral}
\end{figure}

Our approach consists of combining the solution for the Green function obtained by Clinton, Esrick, and Sacks \cite{Clinton-PRB-1985}, with the formula for the vdW interaction obtained by Eberlein and Zietal \cite{Eberlein-PRA-2007}.
Let us start discussing the perturbative analytical solution 
in Ref. \cite{Clinton-PRB-1985}.
For a charge $Q$ located at the position ${\bf r}^{\prime}={\bf r}_{||}^{\prime}+z^\prime\hat{{\bf z}}$
(with $z^\prime>0$ and ${\bf r}_{||}^\prime=x^\prime\hat{{\bf x}}+y^\prime\hat{{\bf y}}$), one has the
Poisson equation
$\boldsymbol{\nabla}^2{\bf \phi\left({\bf r},{\bf r}^{\prime}\right)}=-4\pi Q\delta\left({\bf r}-{\bf r}^{\prime}\right)$,
where the potential $\phi$ can be written as 
$\phi\left({\bf r}, {\bf r}^{\prime}\right)=QG\left({\bf r},{\bf r}^{\prime}\right)$, where
$G\left({\bf r},{\bf r}^{\prime}\right)$ is the Green function of the Laplacian operator.
Let us consider the perturbative analytical solution for $G\left(\textbf{r},\textbf{r}^{\prime}\right)$, 
subjected to the boundary condition $\left.G\left(\textbf{r},\textbf{r}^{\prime}\right)\right|_{z=h({\bf r}_{||})}=0$,
with $h({\bf r}_{||})$ describing a suitable modification $[\text{max}|h({\bf r}_{||})|=a\ll z^\prime]$ 
of a grounded planar conducting surface at $z=0$ (see Fig. \ref{fig:carga-superficie-geral}).
Following Clinton, Esrick, and Sacks \cite{Clinton-PRB-1985}, the approximate solution
of $G$, up to order $h$, is
$G\left(\textbf{r},\textbf{r}^{\prime}\right) \approx G^{(0)}\left(\textbf{r},\textbf{r}^{\prime}\right) + 
G^{(1)}\left(\textbf{r},\textbf{r}^{\prime}\right)$,
with
$G^{(0)}\left(\textbf{r},\textbf{r}^{\prime}\right) ={1}/{|\textbf{r}-\textbf{r}^{\prime}|}
+G^{(0)}_{H}\left(\textbf{r},\textbf{r}^{\prime}\right)$,
where 
$G^{(0)}_{H}\left(\textbf{r},\textbf{r}^{\prime}\right)=-{1}/{[|\textbf{r}_{\parallel}-\textbf{r}_{\parallel}^{\prime}|^{2}
	+\left(z+z^{\prime}\right)^{2}]^{\frac{1}{2}}}$
is the homogeneous part of $G^{(0)}$,
and
\begin{equation}
G^{(1)}(\textbf{r},\textbf{r}^{\prime})=-\int\frac{d^{2}\tilde{\textbf{r}}_{\parallel}}{4\pi} h(\tilde{\textbf{r}}_{\parallel})\left[\frac{\partial G^{(0)}}{\partial\tilde{z}}
(\textbf{r},\tilde{\textbf{r}})\frac{\partial G^{(0)}}{\partial\tilde{z}}
(\tilde{\textbf{r}},\textbf{r}^{\prime})\right]_{\tilde{z}=0}.
\label{eq:sol-G1}
\end{equation}
Note that $G^{(0)}$ is the unperturbed solution, related to a planar surface
at $z=0$ (obeying $\left.G^{(0)}\left(\textbf{r},\textbf{r}^{\prime}\right)\right|_{z=0}=0$).
This approximate solution for $G$ becomes increasingly better as $a/z^\prime\ll 1$.
We also remark that all the perturbative process leading to Eq. \eqref{eq:sol-G1} (and subsequent orders \cite{Clinton-PRB-1985}) is done in terms of the parameter $a/z^\prime$, not requiring any expansion in the arguments of $h$, and, hence, no derivatives of $h$ appear in Eq. \eqref{eq:sol-G1}.
In other words, no demand on the smoothness of the surface ${z=h({\bf r}_{||})}$ is necessary,
so that there is no restriction for the ratio between the mean characteristic distance
along the surface and $z^\prime$, which means that the solution is valid beyond the PFA \cite{Clinton-PRB-1985}.
The classical interaction energy between the point charge and the surface 
depends on the homogeneous part of $G$, namely
$G_H\left(\textbf{r},\textbf{r}^{\prime}\right)=G\left(\textbf{r},\textbf{r}^{\prime}\right)-{1}/{|\textbf{r}-\textbf{r}^{\prime}|}$,
whose perturbative solution is
$G_{H}\left(\textbf{r},\textbf{r}^{\prime}\right) \approx G^{(0)}_{H}\left(\textbf{r},\textbf{r}^{\prime}\right) + G^{(1)}\left(\textbf{r},\textbf{r}^{\prime}\right)$.

Now, we discuss the Eberlein-Zietal method to compute the vdW interaction between a polarizable neutral point particle and a general grounded conducting surface \cite{Eberlein-PRA-2007}. This method consists of mapping the classical interaction energy $U_\text{cla}$,
\begin{equation}
U_\text{cla}(\textbf{r}_0)=\frac{1}{8\pi\epsilon_{0}}\left.(\textbf{d}\cdot\nabla^{\prime})(\textbf{d}\cdot\nabla)G_{H}(\mathbf{r},\mathbf{r}^{\prime})\right|_{\mathbf{r}=\mathbf{r}^{\prime}=\mathbf{r_{0}}},
\label{eq:Eberlein_Zietalclassic}
\end{equation}
which is the interaction energy for a point particle located at ${\bf r}_{0}=x_0\hat{{\bf x}}+y_0\hat{{\bf y}}+z_0\hat{{\bf z}}$
$(z_0>0)$ and with a dipole moment vector $\textbf{d}$, into the van der Waals interaction $U_{\text{vdW}}$,
\begin{equation}
U_{\text{vdW}}(\mathbf{r}_0)=
\frac{1}{8\pi\epsilon_{0}}\displaystyle\sum_{i,j}\langle \hat{d}_i \hat{d}_j\rangle\nabla_i\nabla_j'\left.G_H(\mathbf{r},\mathbf{r'})\right|_{\mathbf{r}=\mathbf{r'}=\mathbf{r}_0},	\label{eq:Eberlein_Zietalquantum}
\end{equation}
where $\hat{d}_i$ $(i,j=\{x,y,z\})$ are the components of the dipole moment operator and
$\langle \hat{d}_i \hat{d}_j\rangle$ is the expectation value of $\hat{d}_i \hat{d}_j$ (see also \cite{Souza-AJP-2013}).
In both equations \eqref{eq:Eberlein_Zietalclassic} and \eqref{eq:Eberlein_Zietalquantum}, the function $G_H$ is the solution of the Laplace equation in the presence of the surface, which contains all the information about its geometry. 

In this paper, we investigate the physical implications of a set of new analytical formulas, developed by us, combining the perturbative solution $G_{H} \approx G^{(0)}_{H} + G^{(1)}$, where $ G^{(1)}$ is given in Eq. \eqref{eq:sol-G1}, with the classical and quantum interactions given by Eqs. \eqref{eq:Eberlein_Zietalclassic} and \eqref{eq:Eberlein_Zietalquantum}. 

%%%%%%%%%%%%%%%%%%%%%%%%%%%%%%%%%%%%%%%%%%%%%%%%%%%%%%%%%%%%%%%%%%%%%%%%%%%%%%%%%%%%%%%%%%
%
\section{The Classical Interaction}
\label{sec-classical-case}

The electrostatic interaction between a polarized particle, at ${\bf r}_{0}=x_0\hat{{\bf x}}+y_0\hat{{\bf y}}+z_0\hat{{\bf z}}$
$(z_0>0)$,
and a grounded conducting rough surface can be written as $U_\text{cla}\approx U^{(0)}_\text{cla} + U^{(1)}_\text{cla}$, where
$ U^{(0)}_\text{cla}(z_0) = -{(d_{x}^{2} +  d_{y}^{2} + 2 d_{z}^{2} )}/{64\pi\epsilon_{0}z_{0}^{3}}$
is the usual interaction energy between a dipole and a grounded conducting plane \cite{Jackson-Electrodynamics-1998}, and $ U^{(1)}_\text{cla} $ is the first-order correction to $U^{(0)}_\text{cla}$ due to the surface corrugation.
Substituting $G^{(1)}$ [given by Eq. \eqref{eq:sol-G1}] in Eq. \eqref{eq:Eberlein_Zietalclassic}, we write $U^{(1)}_\text{cla}$ in terms of $\tilde{h}({\bf q})$, the Fourier representation of $h(\textbf{r}_{\parallel})$, obtaining
\begin{equation}
U^{(1)}_\text{cla}(\mathbf r_0)= -\sum_{i,j}
\frac{d_{i}d_{j}}{64\pi\epsilon_{0}}\int\frac{d^{2}{\bf q}}{(2\pi)^{2}}\tilde{h}({\bf q})e^{i{\bf q}\cdot{\bf r}_{0\parallel}}\mathcal{I}_{ij}(z_{0},{\bf q})
\label{principal}
\end{equation}
where ${\bf q} = q_x\hat{{\bf x}}+q_y\hat{{\bf y}}$, and the functions $\mathcal{I}_{ij}=\mathcal{I}_{ji}$ are written as
$\mathcal{I}_{ij} =\mathcal{K}_{ij}^{(2)} K_{2}\left(z_{0}|{\bf q}|\right)
+\mathcal{K}_{ij}^{(3)} K_{3}\left(z_{0}|{\bf q}|\right)$,
where $K_{2}$ and $K_{3}$ are modified Bessel functions of the second kind, and,
$\mathcal{K}_{ij}^{(s)}=\mathcal{K}_{ji}^{(s)}$ ($s=2,3$)
are given by
$\mathcal{K}_{xx}^{(2)}=-\frac{3}{8}q_{x}^{2}|{\bf q}|^{2}$, 
$\mathcal{K}_{xx}^{(3)}=\frac{3}{8}\frac{|{\bf q}|^{3}}{z_{0}}$,
$\mathcal{K}_{yy}^{(2)}=-\frac{3}{8}q_{y}^{2}|{\bf q}|^{2}$, 
$\mathcal{K}_{yy}^{(3)}=\frac{3}{8}\frac{|{\bf q}|^{3}}{z_{0}}$,
$\mathcal{K}_{zz}^{(2)}=(2+\frac{3}{8}z_{0}^{2}|{\bf q}|^{2})\frac{|{\bf q}|^{2}}{z_{0}^{2}}$,
$\mathcal{K}_{zz}^{(3)}=\frac{1}{4}\frac{|{\bf q}|^{3}}{z_{0}}$,
$\mathcal{K}_{xy}^{(2)}=-\frac{3}{8}q_{x}q_{y}|{\bf q}|^{2}$,
$\mathcal{K}_{xy}^{(3)}=0$,
$\mathcal{K}_{xz}^{(2)}=i\frac{q_{x}|{\bf q}|^{2}}{z_{0}}$,
$\mathcal{K}_{xz}^{(3)}=-\frac{3i}{8}q_{x}|{\bf q}|^{3}$,
$\mathcal{K}_{yz}^{(2)}=i\frac{q_{y}|{\bf q}|^{2}}{z_{0}}$,
$\mathcal{K}_{yz}^{(3)}=-\frac{3i}{8}q_{y}|{\bf q}|^{3}$.
It is worth to point out that the roughness correction $U^{(1)}_{\text{cla}}$, given by Eq. \eqref{principal}, is valid for any roughness profile function $h$ with $\max(|h|)\ll z_0$, with no constraint on the smoothness of the surface,
which means that we can investigate configurations beyond the scope of the PFA.

\begin{figure}[h]
	\centering
	\epsfig{file=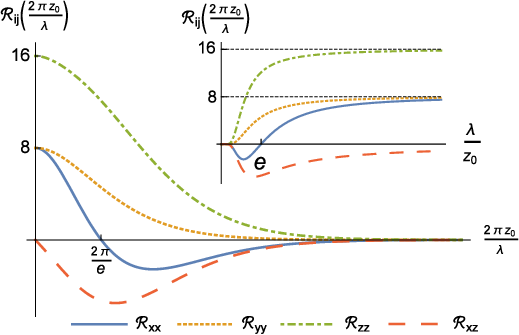,  width=0.8 \linewidth}
	\caption{
		The behavior of ${\cal R}_{ij}(2\pi z_0/\lambda)$ versus $2\pi z_0/\lambda$.
		The function $\mathcal{R}_{xx}$ (solid line) changes its sign at $2\pi z_0/\lambda \approx 2\pi/e$, where $e$ is the Euler number.
		In the inset we also show the behavior of ${\cal R}_{ij}(2\pi z_0/\lambda)$ versus $\lambda/z_0$.
		Note that the function $\mathcal{R}_{xx}$ changes its sign at $\lambda/z_0 \approx e$.
	}
	\label{fig:response-functions}
\end{figure}

\begin{figure}[h]
	\centering
	\epsfig{file=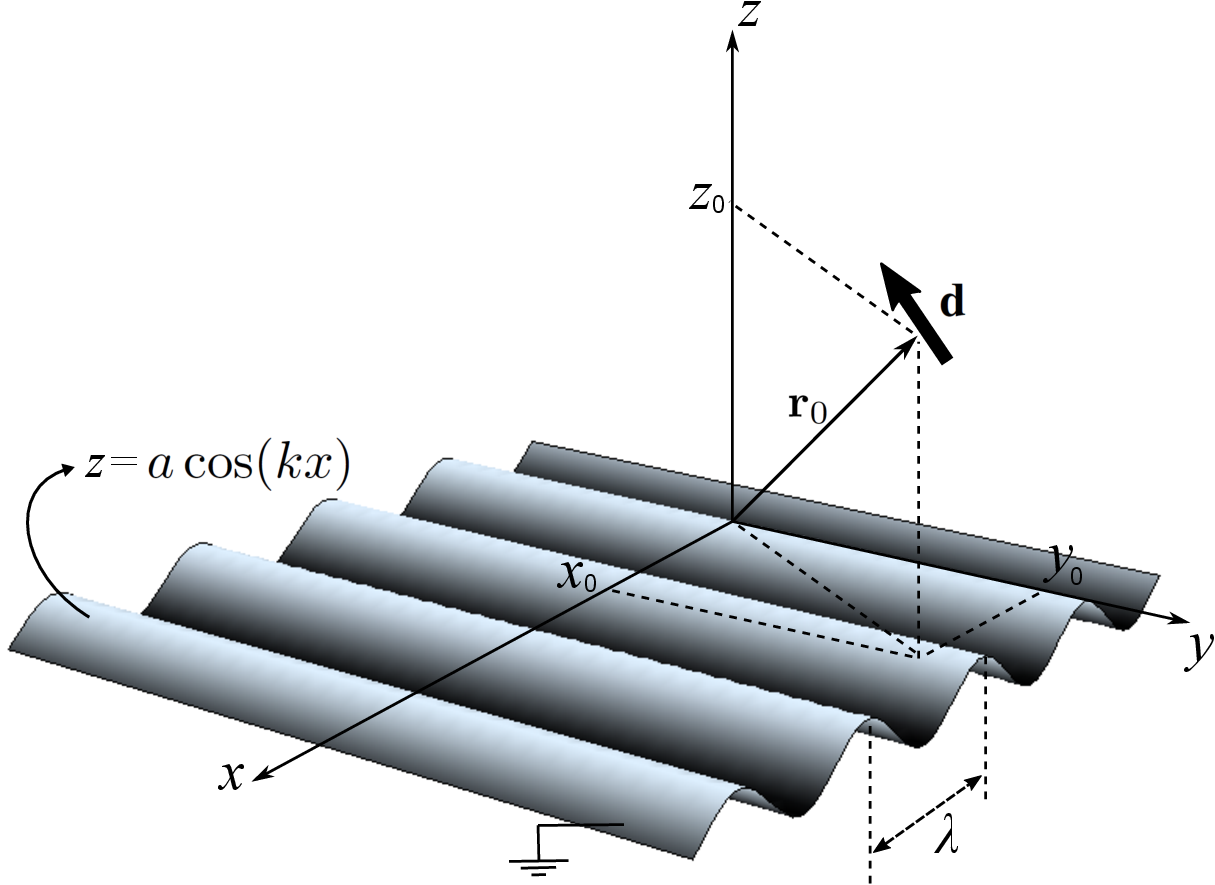,  width=0.8 \linewidth}
	\caption{
		Illustration of a particle with a dipole moment $\textbf{d}$, located at ${\bf r}_{0}=x_0\hat{{\bf x}}+y_0\hat{{\bf y}}+z_0\hat{{\bf z}}$ (with $z_0>0$), interacting with a sinusoidal grounded conducting corrugated surface, whose corrugation profile is described by $z= a\cos(kx)$, with $a \ll z_0$.
	}
	\label{fig:dipolo-superficie-senoidal}
\end{figure}

Let us investigate the case of a sinusoidal corrugated surface with amplitude $a$ and corrugation period $\lambda$, which is described by
$h(x)=a\cos(k x)$,
where  $k = 2\pi/\lambda$ and $a\ll z_0$ (see Fig. \ref{fig:dipolo-superficie-senoidal}).
Then, the integrals in Eq. \eqref{principal} can be analytically solved \cite{Gradshteyn-Table-2007,Watson-BesselFunctions-1944}, resulting in
\begin{equation}
U^{(1)}_\text{cla}=-\frac{3a}{512\pi\epsilon_{0}z_{0}^{4}}A(d_i d_j,kz_{0})
\cos[kx_{0}-\delta(d_i d_j,kz_{0})],
\label{eq:potential-energy}
\end{equation}
where $\delta$ is  a nontrivial phase function defined by
\begin{equation}
\sin(\delta)=\frac{B(d_i d_j,kz_{0})}{A(d_i d_j,kz_{0})},\;\;\;\cos(\delta)=\frac{C(d_i d_j,kz_{0})}{A(d_i d_j,kz_{0})}, 
\label{eq:delta}
\end{equation}
with $A=\sqrt{B^{2}+C^{2}}$,
\begin{eqnarray}
B &=& -2d_{x}d_{z}{\cal R}_{xz}(kz_0), \label{eq:B}\\
C &=& d_{x}^{2}{\cal R}_{xx}(kz_0)+d_{y}^{2}{\cal R}_{yy}(kz_0)+d_{z}^{2}{\cal R}_{zz}(kz_0),
\label{eq:C}
\end{eqnarray}
and the functions ${\cal R}_{ij}$ are defined by:
${\cal R}_{xx}(u)=u^{3}K_{3}(u)-u^{4}K_{2}(u)$,
${\cal R}_{yy}(u)=u^{3}K_{3}(u)$,
${\cal R}_{zz}(u)=(u^{4}+\frac{16}{3}u^{2})K_{2}(u)+\frac{2}{3}u^{3}K_{3}(u)$,
${\cal R}_{xz}(u)=\frac{8}{3}u^{3}K_{2}(u)-u^{4}K_{3}(u)$.
We highlight that just one of them, namely ${\cal R}_{xx}$, changes its sign,
which happens at the point $kz_0\approx 2\pi/e$, with this value corresponding
to $\lambda/z_0\approx e$, where $e$ is the Euler number (see Fig. \ref{fig:response-functions}).
This change of sign of ${\cal R}_{xx}$ has a fundamental role on the results obtained in the present paper, because it affects the value of the phase function $\delta$ in a nontrivial way, as we will see soon.
In order to simplify the discussions, hereafter we make our analysis in terms of the parameter $\lambda/z_0$.

\begin{figure}[h]
\centering 
\epsfig{file=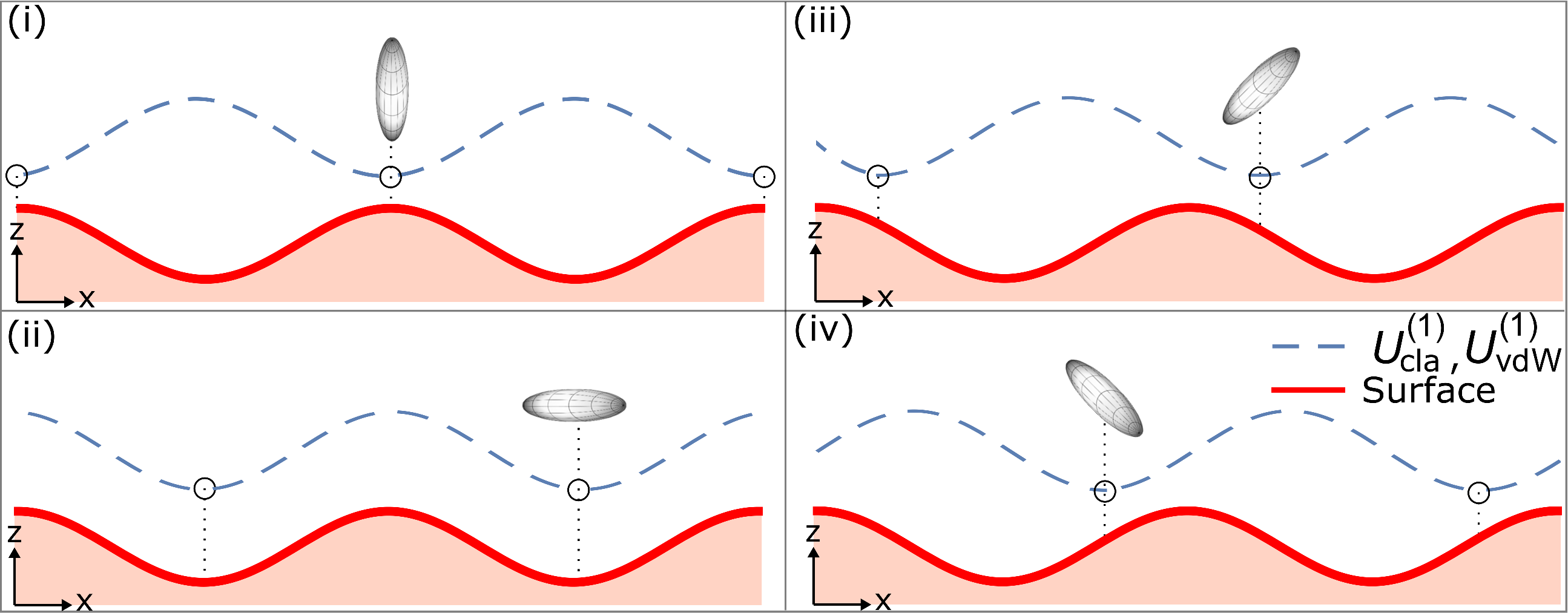,  width=1.0 \linewidth}  
\caption{Illustration of a neutral anisotropic particle (represented by an ellipsoid), fixed at $z=z_0$,
and interacting with a rough surface (solid lines).
The stable equilibrium points (indicated by the circles) of $U^{(1)}_\text{cla}$ [Eq. \eqref{eq:potential-energy}]
(dashed lines) can be over the corrugation peaks [peak regime, (i)], valleys [valley regime, (ii)], or over points between a peak and a valley [intermediate regime, shown in (iii) and (iv)]. 
We highlight the 
different equilibrium
points in (iii) and (iv), which is related to the change $d_z\to-d_z$
(or $d_x\to-d_x$) in the orientation of the particle.
The largest axis of the ellipsoid represents the direction of the dipole moment $\textbf{d}$ of the particle in each situation.	
The same figure also illustrates
the stable equilibrium points of the vdW energy $U^{(1)}_\text{vdW}$ [Eq. \eqref{eq:potential-energy-CP}]. 
In this case,
the largest axis of the ellipsoid represents the direction in which
the tensor
$\langle \hat{d}_{i}\hat{d}_{j}\rangle$
diagonalized has its higher value.  
}
\label{fig:regimes}
\end{figure}

\begin{figure}
\centering  
\subfigure[]{\label{fig:regiao-classico}\epsfig{file=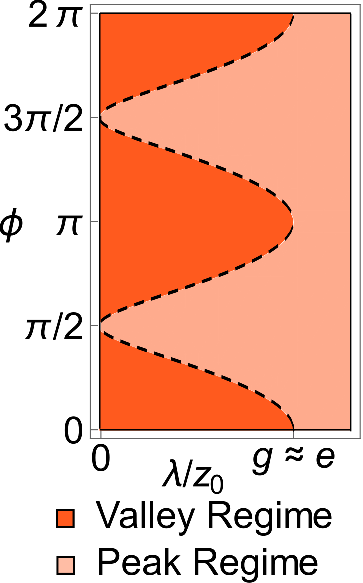, width=0.2 \linewidth}}
\hspace{3mm}
\subfigure[]{\label{fig:max-min-classico}\epsfig{file=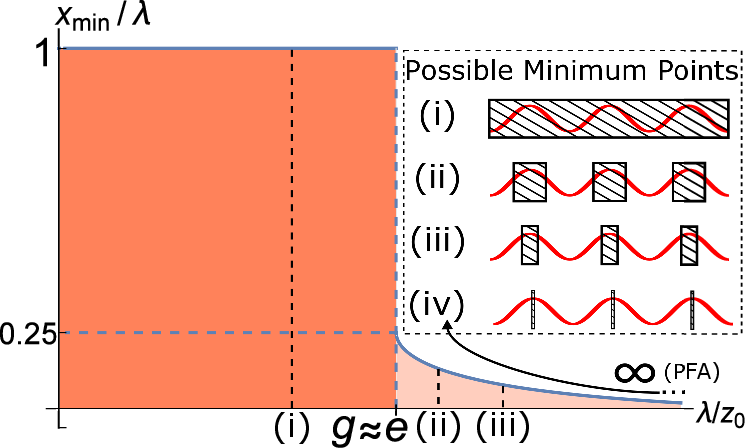, width=0.6 \linewidth}}
\caption{
(a) Configuration space representing $\phi$ (vertical axis) and $\lambda/z_0$ (horizontal axis).
It shows the behavior of $x_{\text{min}}$ of $U^{(1)}_\text{cla}$ for a particle with magnitude of  dipole moment $|\textbf{d}|$. 
The dark region represents $C(\theta=\pi/2)<0$, and corresponds to the valley regime.
The lighter region represents $C(\theta=\pi/2) > 0$, and corresponds to the peak regime.
The border between these two regions (dashed line) corresponds to the situations where $C(\theta=\pi/2) = 0$, which means that 
$ U_{\text{cla}}^{(1)} = 0 $ and that the lateral force vanishes.
(b) The possible values of $x_{\text{min}}/\lambda$ (in the interval $0\leq x_{\text{min}}/\lambda\leq 1$), versus $\lambda/z_0$, considering
all orientations of the particle.
The different shades, dark and lighter regions, represent distinct behaviors, namely, in dark region
$0\leq x_{\text{min}}/\lambda\leq 1$, whereas in lighter one $0\leq x_{\text{min}}/\lambda\leq\beta$,
with the values of $\beta$ given by the upper curve delimiting the lighter region,
from which one can see that $0<\beta(\lambda/z_0)<0.25$ and $\beta(\lambda/z_0\to\infty)=0$.
The inset (a)-(i) considers the periodicity of $U^{(1)}_{\text{cla}}$ and illustrates (with hatched areas) that $x_{\text{min}}$ can assume any value. 
The insets (a)-(ii) and (a)-(iii), in a similar manner, illustrate that the values of $x_{\text{min}}$ can be located only around the peaks.
Inset (a)-(iv) (the PFA limit) illustrates that, when $\lambda/z_0\to\infty$, the minimum points are located over the peaks.
Figures similar to (a) and (b) also appear in the vdW interaction,
but with the separation between the dark and lighter regions occurring for $\lambda/z_0 =g< e$.
}
\label{fig:regiao}
\end{figure}

\begin{figure}[h!]
	\centering  
	\subfigure[$\phi=0$ and $\lambda/z_0<e$]{\label{fig:interm-menor-phi-0}\epsfig{file=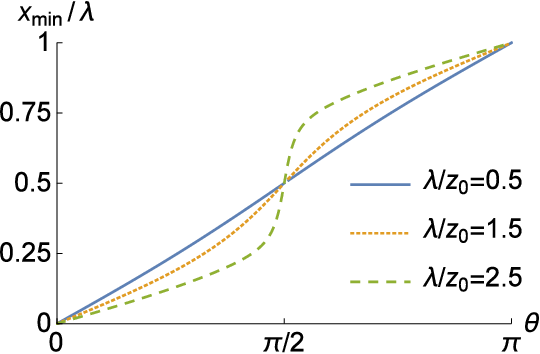, width=0.45 \linewidth}}
	\hspace{5mm}
	\subfigure[$\phi=0$ and $\lambda/z_0>e$]{\label{fig:interm-maior-phi-0}\epsfig{file=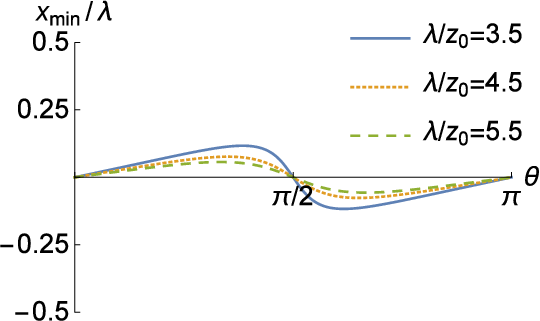, width=0.45 \linewidth}}
	\subfigure[$\phi=\pi/4$ and $\lambda/z_0<1.74$]{\label{fig:interm-menor-phi-45}\epsfig{file=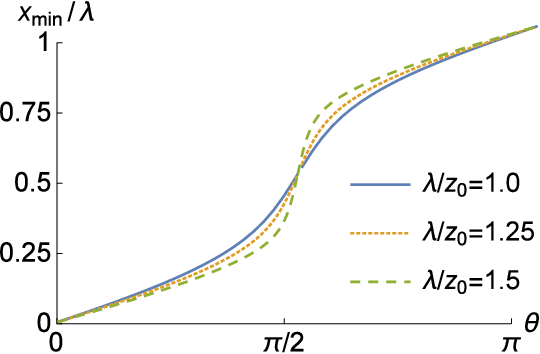, width=0.45 \linewidth}}
	\hspace{5mm}
	\subfigure[$\phi=\pi/4$ and $\lambda/z_0>1.74$]{\label{fig:interm-maior-phi-45}\epsfig{file=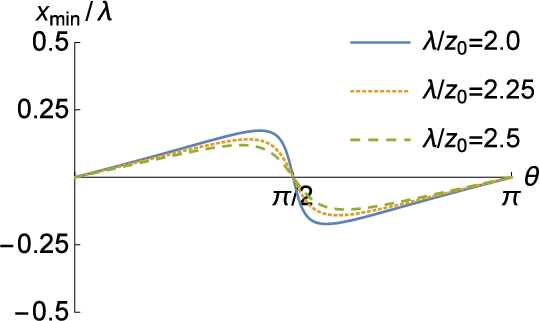, width=0.45 \linewidth}}
	\caption{
		Relation $ x_{\text{min}}/\lambda \times \theta$ for a particle with a dipole moment $\textbf{d}$, with $\phi$ fixed. 
		For a given value of $\phi$, the shape of this relation depends if the value of $\lambda/z_0$ is below [Figs. (a) and (c)] or above [Figs. (b) and (d)] that shown by the dashed border line in Fig. 2(a) of the manuscript.
		For instance, when $\phi = 0$, the value of $\lambda/z_0$ indicated by the dashed border line in Fig. 2(a) of the manuscript is $\lambda/z_0 \approx e$, so that Figs. (a) and (b) show $ x_{\text{min}}/\lambda \times \theta$ for $\lambda/z_0 < e$, and $\lambda/z_0 > e$, respectively.
		When $\phi = \pi/4$, the correspondent value of $\lambda/z_0$ belonging to the dashed border line in Fig. 2(a) of the manuscript is $\lambda/z_0 \approx 1.74$, so that Figs. (c) and (d) show $ x_{\text{min}}/\lambda \times \theta$ for $\lambda/z_0 < 1.74$, and $\lambda/z_0 > 1.74$, respectively.
	}
	\label{fig:regime-interm-classico}
\end{figure}

From Eq. \eqref{eq:potential-energy},
considering the particle fixed
at $z=z_0$, we can see that the stable equilibrium points of $U^{(1)}_\text{cla}$ can be over the corrugation peaks (when $\delta=0$), valleys ($\delta=\pi$), or over points between a peak and a valley ($\delta \neq 0,\pi$).
Such behaviors are named here as peak, valley and intermediate regimes, respectively, and can 
be visualized in Fig. \ref{fig:regimes}.
From Eqs. \eqref{eq:delta}-\eqref{eq:C}, we can see that when $d_x=0$ one has only the peak regime, for $d_z=0$ we can have peak or valley regimes, whereas for other dipole orientations we have intermediate regimes.
We will now investigate each one of these possibilities separately.
For $d_x=0$, from Eqs. \eqref{eq:B} and \eqref{eq:C}, we have the functions $B=0$ and $C>0$ (since ${\cal R}_{xx}$, the only function ${\cal R}_{ij}$ that can be negative, can not contribute in this case), resulting in $\delta=0$ [peak regime, Fig. \ref{fig:regimes}(i)] for any value of $d_y$ and $d_z$.
On the other hand, for $d_z=0$, we have $B=0$ again, but now $C$ can be negative.
If we write the components of the dipole vector in spherical coordinates, namely 
$d_x = |\textbf{d}| \sin\theta \cos\phi $, $d_y = |\textbf{d}| \sin\theta \sin\phi $ and $d_z = |\textbf{d}| \cos\theta $, we obtain that the valley regime [Fig. \ref{fig:regimes}(ii)] occurs 
for $\theta=\pi/2$ and $\tan^2(\phi)<-\mathcal{R}_{xx}/\mathcal{R}_{yy}$ (dark region in Fig. \ref{fig:regiao-classico}), whereas the peak regime occurs for $\tan^2(\phi)>-\mathcal{R}_{xx}/\mathcal{R}_{yy}$ (lighter region in Fig. \ref{fig:regiao-classico}).
The border between these two regions (dashed line in Fig. \ref{fig:regiao-classico}) corresponds to the situations where 
$ U^{(1)}_{\text{cla}} = 0 $, which means that the lateral force vanishes.
Note that the highest value of $\lambda/z_0$ which belongs to this dashed border occurs when $\phi = 0,\pi$ 
(dipole oriented in $x$-direction), and here, it happens at $\lambda/z_0 \approx e$, so that above it there is no value of $\phi$ making possible the valley regime.
From Eqs. \eqref{eq:B} and the formula for ${\cal R}_{xz}$, one can see that when $d_x\neq0$ and $d_z\neq0$ we have the function $B \neq 0$, resulting in the intermediate regime [Fig. \ref{fig:regimes}(iii) and Fig. \ref{fig:regimes}(iv)].
In this case, the values $x_{\text{min}}$ (minimum values 
of $U^{(1)}_\text{cla}$) can be computed analytically by solving the equation $kx_{\text{min}}-\delta(d_i d_j,kz_{0}) = 2n\pi$, with $ n \in \mathbb{Z} $, with the phase function $\delta$ obtained from Eqs. \eqref{eq:delta}.
When we consider general orientations $(\phi, \theta)$ of the particle,
the solution of this equation
reveals other two more general behaviors of $U^{(1)}_\text{cla}$,
as shown in Fig. \ref{fig:max-min-classico}.
The dark region in Fig. \ref{fig:max-min-classico} represents that,
with an appropriate choice of $(\phi,\theta)$, $x_{\text{min}}/\lambda$ can have any value in the interval $[0,1]$,
with this behavior occurring for $0<\lambda/z_0 \lesssim e$. The inset in Fig. \ref{fig:max-min-classico}
considers the periodicity of $U^{(1)}_{\text{cla}}$, and illustrates in (i) that the dark region implies that
an appropriate choice of $(\phi,\theta)$ can give any value to $x_{\text{min}}$.
This behavior can also be seen in Figs. \ref{fig:interm-menor-phi-0}, \ref{fig:interm-menor-phi-45}
and \ref{fig:interm-maior-phi-45}.
The lighter region in Fig. \ref{fig:max-min-classico} means that the orientations $(\phi, \theta)$ only can produce values of $x_{\text{min}}$ 
located around the peaks, as shown in insets (ii) and (iii).
This behavior can also be seen in Fig. \ref{fig:interm-maior-phi-0}.
One can also see in Fig. \ref{fig:max-min-classico} that, as $\lambda/z_0$ increases, the region where the possible $x_{\text{min}}$ can be found decreases, so that, when $\lambda/z_0\to\infty$ (PFA), we have $x_{\text{min}}/\lambda\to n$, which means that the minimum points are only over the peaks, independently of $(\phi,\theta)$ [see Fig. \ref{fig:max-min-classico}, inset (iv)].
This can be understood by taking the limit $\lambda/z_0 \to \infty$ in our Eq. \eqref{eq:potential-energy}, leading to $U^{(1)}_\text{cla}\to-a\cos(kx_{0})({3}/{64\pi\epsilon_{0}z_{0}^{4}})(d_{x}^{2}+d_{y}^{2}+2d_{z}^{2})=-h(x_0)dU^{(0)}_\text{cla}(z_0)/dz_0 $,
which is characteristic of the PFA \cite{Clinton-PRB-1985}.
In other words, the PFA just sees the peak regime.

%%%%%%%%%%%%%%%%%%%%%%%%%%%%%%%%%%%%%%%%%%%%%%%%%%%%%%%%%%%%%%%%%%%%%%%%%%%%%%%%%%%%%%%%%%
\section{Van der Waals interaction}
\label{sec-our-caimir-polder}

Our formula for the vdW interaction between a polarizable particle and a grounded conducting rough surface can be written as
$U_\text{vdW}\approx U^{(0)}_{\text{vdW}} + U^{(1)}_{\text{vdW}}$, where $ U^{(0)}_{\text{vdW}}(z_0) = -{(\langle \hat{d}_{x}^{2}\rangle+ \langle\hat{d}_{y}^{2} \rangle + 2\langle \hat{d}_{z}^{2} \rangle )}/{64\pi\epsilon_{0}z_{0}^{3}}$ is the vdW potential for the case of a grounded conducting plane \cite{Lennard-Jones-TransFarSoc-1932}, and $ U^{(1)}_{\text{vdW}} $ (the first-order correction of $U^{(0)}_{\text{vdW}}$ due to the surface corrugation) is obtained by substituting $G^{(1)}$, given by Eq. \eqref{eq:sol-G1}, in Eq. \eqref{eq:Eberlein_Zietalquantum}, obtaining
\begin{equation}
	U^{(1)}_{\text{vdW}}(\mathbf r_0)= -\sum_{i,j}\frac{\langle  \hat{d}_{i}\hat{d}_{j} \rangle}{64\pi\epsilon_{0}}
	\int\frac{d^{2}{\bf q}}{(2\pi)^{2}}\tilde{h}({\bf q})e^{i{\bf q}\cdot{\bf r}_{0\parallel}}\mathcal{I}_{ij}(z_{0},{\bf q}),
	\label{principal-quantico}
\end{equation}
where 
$
\langle \hat{d}_{i} \hat{d}_{j}\rangle = \frac{\hbar}{\pi}\int_{0}^{\infty}d\xi\,\alpha_{ij}\left(i\xi\right),
$
with $\alpha_{ij}$ the components of the polarizability
tensor \cite{Buhmann-DispersionForces-I,Buhmann-DispersionForces-II}.
We have that this formula \eqref{principal-quantico} is valid for any roughness profile function $h$ with $\max(|h|)\ll z_0$, with no  constraint on the smoothness of the surface, therefore valid beyond the PFA. 
For the case of a sinusoidal corrugated surface, Eq. \eqref{principal-quantico} leads to
\begin{equation}
	U^{(1)}_{\text{vdW}}=-\frac{3a A(\langle\hat{d}_i \hat{d}_j\rangle,kz_{0})}{512\pi\epsilon_{0}z_{0}^{4}}
	\cos[kx_{0}-\delta(\langle\hat{d}_i \hat{d}_j\rangle,kz_{0})],
	\label{eq:potential-energy-CP}
\end{equation}
where $A(\langle\hat{d}_i \hat{d}_j\rangle,kz_{0})$ and $\delta(\langle\hat{d}_i \hat{d}_j\rangle,kz_{0})$ 
are obtained by replacing 
${d}_i {d}_j \rightarrow \langle\hat{d}_i \hat{d}_j\rangle$ in Eqs. \eqref{eq:delta}-\eqref{eq:C}.
Taking the limit 
$\frac{\lambda}{z_0} \to \infty$
one obtains $U^{(1)}_{\text{vdW}}\to-a\cos(kx_{0})({3}/{64\pi\epsilon_{0}z_{0}^{4}})(\langle d_{x}^{2} \rangle + \langle d_{y}^{2} \rangle
+2 \langle d_{z}^{2}\rangle)=-h(x_0)dU^{(0)}_{\text{vdW}}(z_0)/dz_0$.
This result is characteristic of the PFA \cite{Dalvit-JPA-2008}, and one can see that this limit eliminates the presence of the phase function $\delta$, so that valley and intermediate regimes can not be predicted if this approximation is used.

For isotropic particles, 
$\langle\hat{d}_x \hat{d}_z\rangle = 0$ and 
$\langle\hat{d}^{2}_x\rangle=\langle\hat{d}^{2}_y\rangle=\langle\hat{d}^{2}_z\rangle$,
which leads to $B(\langle\hat{d}_i \hat{d}_j\rangle,kz_{0})=0$ and $C(\langle\hat{d}_i \hat{d}_j\rangle,kz_{0})>0$.
Then, from Eq. \eqref{eq:delta}, we obtain that $\delta=0$, which implies that only the peak regime occurs, and our Eq. \eqref{eq:potential-energy-CP} recovers that for the  vdW interaction obtained in Ref. \cite{Dalvit-PRL-2008}.
For an anisotropic particle oriented with its principal axes coinciding with $xyz$, 
according to Eq. \eqref{eq:C} $B=0$, so that only two situations are possible: $\delta=0$ [peak regime, Fig. \ref{fig:regimes}(i)] or
$\delta=\pi$ [valley regime,
Fig. \ref{fig:regimes}(ii)]. 
This reveals behaviors of 
$U^{(1)}_\text{vdW}$ similar to those
shown in Fig. \ref{fig:regiao-classico},
but with the separation between the dark and lighter regions occurring here for 
$\lambda/z_0 =g< e$.
This comes from the distinctions between the classical and quantum versions of Eqs. \eqref{eq:B} and \eqref{eq:C}.
In both, $B=0$, but in the classical case
any $d_i$ can be set to null, so that we can eliminate, arbitrarily, 
${\cal R}_{xx}$, ${\cal R}_{yy}$ or ${\cal R}_{zz}$ factors in Eq. \eqref{eq:C}, whereas in the quantum case one has
$\langle\hat{d}^{2}_x\rangle$,
$\langle\hat{d}^{2}_y\rangle$
and $\langle\hat{d}^{2}_z\rangle$
non-null, independently of the orientation of the particle, implying that all terms ${\cal R}_{ii}$
always contribute, resulting in $\lambda/z_0 < e$.
In contrast, when the particle is oriented in such a way that
its principal axes do not coincide with $xyz$, $B \neq 0$, so that Eq. \eqref{eq:delta} leads to the presence of the phase function $\delta$ in such a way that $\delta \neq 0$ and $\delta \neq \pi$. This means that the intermediate regime occurs [see Fig. \ref{fig:regimes}(iii)-(iv)].

%%%%%%%%%%%%%%%%%%%%%%%%%%%%%%%%%%%%%%%%%%%%%%%%%%%%%%%%%%%%%%%%%%%%%%%%%%%%%%%%%%%%%%%%%%
\section{Applications}
\label{sec-our-casimir-polder}

\begin{figure}
	\centering  
	\subfigure[Sinusoidal]{\label{fig:regimes-seno}\epsfig{file=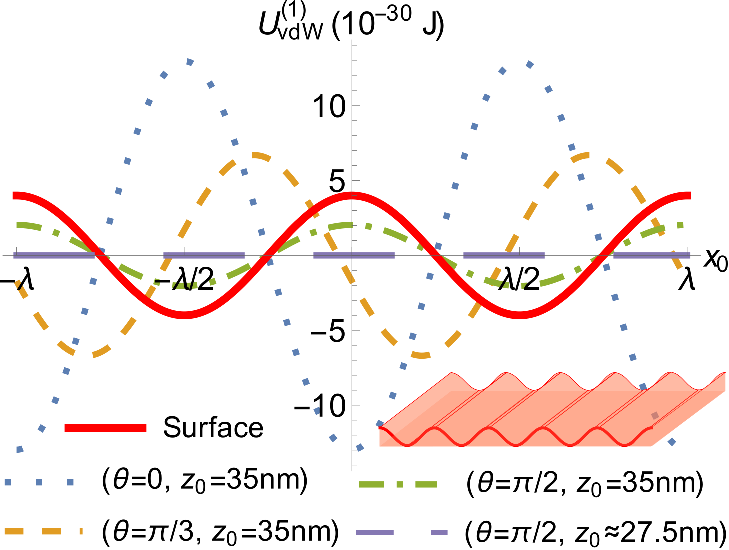, width=0.45 \linewidth}}
	\hspace{0mm}
	\subfigure[Rectangular]{\label{fig:regimes-retang}\epsfig{file=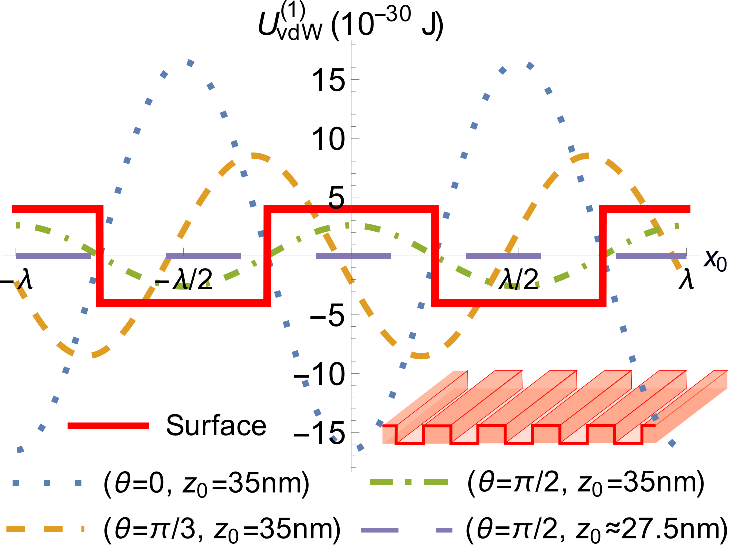, width=0.45 \linewidth}}
	\hspace{0mm}
	\subfigure[Trapezoidal]{\label{fig:regimes-trapezio}\epsfig{file=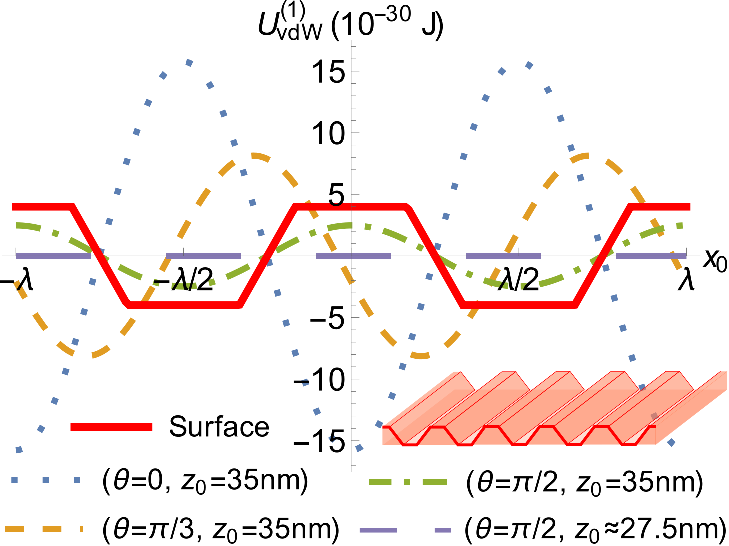, width=0.45 \linewidth}}
	\subfigure[Nonperiodic/Rectangular]{\label{fig:regimes-dois-morros}\epsfig{file=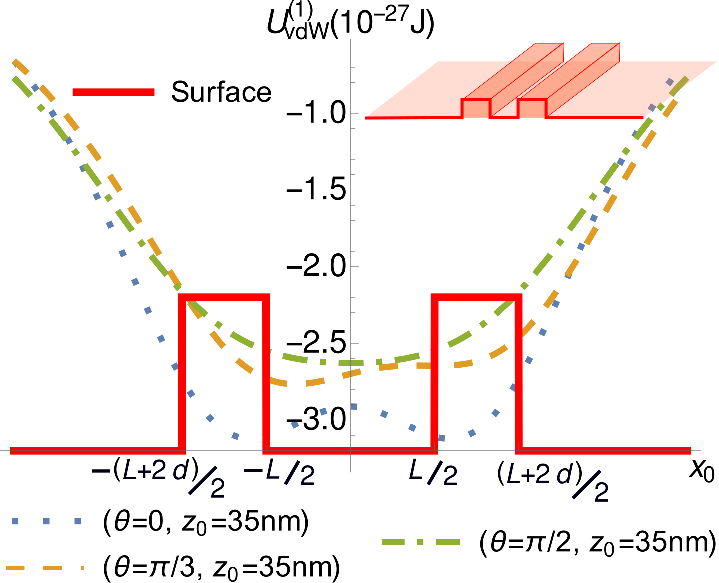, width=0.45 \linewidth}}
	\caption{
		Behavior of $U^{(1)}_\text{vdW}$ versus $x_0$, for a $\text{CO}_2$ molecule interacting with nanogratings
		(illustrated in the insets, and whose profiles are represented by the solid lines). 
		The molecule is fixed at values of $z_0$ within the vdW regime.
		We consider the nanogratings with a depth of $4$nm,
		$\lambda=20$nm, $L=20$nm and $d=10$nm. 
		The dotted lines ($\theta=0$ and $z_0=35$nm) show
		the peak regimes [here meaning that the minimal points of
		$U^{(1)}_\text{vdW}$ are over $x_0=n\lambda$ $(n\in\mathbb{Z})$
		in Figs. \ref{fig:regimes-seno}-\ref{fig:regimes-trapezio},
		and  $x_0\approx\pm(L+d)/2$ in Fig. \ref{fig:regimes-dois-morros}].
		The dashed lines ($\theta=\pi/2$ and $z_0=35$nm)
		show the valley regimes [here meaning that the minimal points
		are over $x_0=(2n+1)\lambda/2$ $(n\in\mathbb{Z})$
		in Figs. \ref{fig:regimes-seno}-\ref{fig:regimes-trapezio},
		and  $x_0= 0$ in Fig. \ref{fig:regimes-dois-morros}].
		The dot-dashed lines ($\theta=\pi/3$ and $z_0=35$nm)
		show the intermediate regimes.
		The long-dashed lines ($\theta=\pi/2$ and $z_0\approx27.5$nm) show, in Figs.
		\ref{fig:regimes-seno}-\ref{fig:regimes-trapezio}, a situation where 
		$U^{(1)}_\text{vdW}=0$ (lateral
		vdW force approximately null).
	}
	\label{fig:gratings}
\end{figure}

\begin{figure}[h]
	\centering  
	\epsfig{file=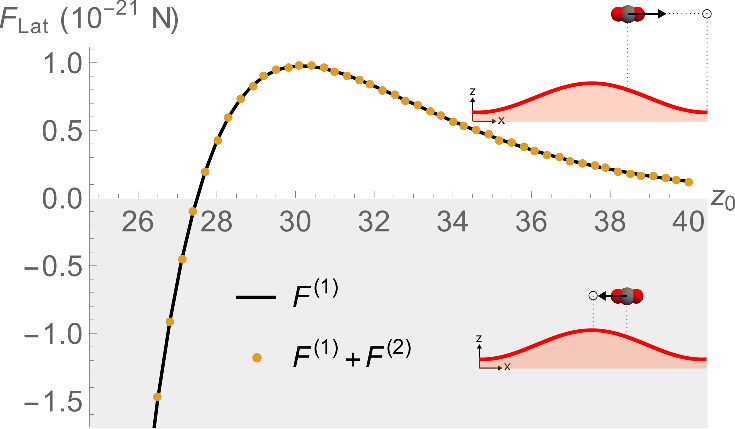, width=0.8 \linewidth}
	\caption{
		Behavior of the lateral vdW force versus $z_0$, for a $\text{CO}_2$ molecule interacting with a sinusoidal nanograting. 
		We consider the nanograting with $a=2$nm and $\lambda=20$nm.
		The molecule is oriented in $x$-direction and fixed at $x_0=2.5$nm.
		The solid line represents the lateral vdW force computed up to first perturbative order $(F^{(1)})$, whereas the dots represent this force computed up to second perturbative order $(F^{(1)}+F^{(2)})$.
		Note that, the second order correction just implies in a minor correction to the first order lateral vdW force, since the dots practically coincides with the solid line.
		We remark that when the force is negative (gray region) the molecule is attracted to the nearest corrugation peak (peak regime), as illustrated in the inset shown in the gray region. 
		When the force is positive (white region) the molecule is attracted to the nearest corrugation valley (valley regime), as illustrated in the inset shown in the white region.
		We highlight that at $z_0 \approx 27.5$, the lateral vdW force is $\approx 0$.
	}
	\label{fig:segunda-ordem}
\end{figure}

Since our Eq. \eqref{principal-quantico} 
[and also Eq. \eqref{principal}] is applicable to any other surface profile, the valley and intermediate regimes, discussed until now for a sinusoidal surface, can also be investigated for gratings commonly considered in Casimir and CP/vdW interactions
\cite{Dalvit-PRL-2008,Lussange-PRA-2012,Bennett-PRA-2015,Zhao-PRA-2008,Lonij-PRA-2009,Lambrecht-PRL-2008}.
For instance, we apply our formulas to the periodic profiles shown in Figs. \ref{fig:regimes-seno}-\ref{fig:regimes-trapezio}, and to the non-periodic one shown in Fig. \ref{fig:regimes-dois-morros}.
In these applications, we consider a $\text{CO}_2$ molecule interacting with nanogratings, 
since the CP/vdW interaction of a such molecule with surfaces has attracted growing interest in technological applications \cite{Antezza-PRB-2020,Huang-JPCL-2015,Titus-ChemRev-2014,Thiyam-CSAPEA-2015,Thiyam-PRA-2015}.
We also consider the dimensions of the nanogratings taking into account the recent advances in fabrication of nanostructures
\cite{Assenbergh-Nanofabrication-2018,
	Li-Nanoscale-2020,Manoccio-Micromachines-2020}, and
the $\text{CO}_2$ polarizability properties as given in Ref. \cite{Thiyam-PRA-2015}. 
In all these situations, one can see the presence of the valley and intermediate regimes, which reveals the 
generality of our results.
Moreover, in these situations, considering the second perturbative order, we obtain that it just implies a minor correction to the first order lateral vdW force (see Fig. \ref{fig:segunda-ordem}), preserving the existence of the new regimes predicted here.

As another application of our results, let us consider 
a particle with mass $m$, fixed at $z_{0}$, put initially at rest in a position 
around a stable equilibrium point of $U^{(1)}_{\text{vdW}}$ (or
$U^{(1)}_{\text{cla}}$, for the classical case).
We obtain that this particle, under the action of the lateral force due to a sinusoidal surface, describes a harmonic oscillation, with frequency  $f=\sqrt{\frac{3aA\left(\lambda,z_{0},\phi,\theta\right)}{512\pi\lambda^{2}z_{0}^{4}\epsilon_{0}m}}$.
For example, a $\text{CO}_2$ molecule,
in the situations shown
in Fig. \ref{fig:regimes-seno},
oscillates with frequency $f\approx 21.06 \text{kHz}$,
$f\approx 15.13 \text{kHz}$, and $f\approx 8.35 \text{kHz}$, in the peak (dotted line),  intermediate  (dashed line), and valley (dot-dashed line) regimes, respectively.
Such frequencies could be detectable by trapping the particle and measuring
the relative shift in the original trap frequency (in the absence of the sinusoidal surface) \cite{Buhmann-DispersionForces-I,Dalvit-PRL-2008}.
We highlight that, in the situation described by the long-dashed line
in Fig. \ref{fig:regimes-seno} (null lateral vdW force), the original trap frequency does not change (even in the presence of a corrugated surface).

As discussed in Ref. \cite{Eberlein-PRA-2007}, 
the reflectivity of a surface is not perfect, 
so that the vdW interaction with a real surface differs by
a numerical factor from the one calculated considering an ideal surface.
In this way, when considering nonideal conductors, 
the values of the vdW energies in Fig. \ref{fig:gratings} need some adjustment, which can be estimated as follows.
Taking into account, for instance, plane surfaces of gold or copper (and their spectral properties given in Ref. \cite{Durand-Tese-2013}), the vdW interaction energies between a $\text{CO}_2$ molecule and these surfaces are, approximately, 0.21 of the correspondent vdW energy when one considers a plane ideal conductor.
Thus, we expect that, in the presence of these nonideal materials, the correspondent energies $U^{(1)}_\text{vdW}$ differ from the results calculated here by a factor of the same order, but preserving the new valley and intermediate regimes. 	

%%%%%%%%%%%%%%%%%%%%%%%%%%%%%%%%%%%%%%%%%%%%%%%%%%%%%%%%%%%%%%%%%%%%%%%%%%%%%%%%%%%%%%%%%%
%
\section{Final remarks}
\label{sec-final}

Nontrivial behaviors of the CP/vdW forces can be predicted when considering anisotropic particles (for instance, the repulsion between a particle and a plate with a hole
\cite{Levin-PRL-2010}) or calculations beyond the PFA 
(which, for example, show that an atom above the plateau of a grooved
plate feels a lateral force, which is not predicted by the PFA \cite{Dalvit-PRL-2008}). 
Here, considering both, anisotropic particles and analytical calculations valid beyond the PFA, we predict new nontrivial behaviors of the lateral vdW force, revealing that, under the action of this force, the particle can be attracted not only toward the corrugation peaks (as found in the literature), but also to the nearest valley, or to an intermediate point between a peak and a valley.
We also show that in the configurations of transition between the peak and valley regimes the lateral vdW force vanishes, even in the presence of a corrugated surface.
Moreover, we show that these new valley and intermediate regimes occur in general, for periodic and nonperiodic corrugated surfaces, and also that similar effects occur for the
classical interaction between a corrugated surface and a particle presenting a permanent electric dipole moment.
These new regimes of lateral force can be relevant for a higher degree of control of the interaction between neutral anisotropic particles and corrugated surfaces in both classical (macroscopic dipoles, ferroelectric nanoparticles or polar molecules) and quantum physics (anisotropic molecules, ellipsoidal nanoparticles), with experimental verifications feasible in both domains. 

%%%%%%%%%%%%%%%%%%%%%%%%%%%%%%%%%%%%%%%%%%%%%%%%%%%%%%%%%%%%%%%%%%%%%%%%%%%%%%%%%%%%%%%%%%%
\begin{acknowledgments}
	The authors thank Alessandra N. Braga, Alexandre Costa, Amanda E. da Silva, Andreson L. C. Rego, Carlos Farina,  Danilo C. Pedrelli, Jeferson D. L. Silva, Nuno M. R. Peres, Paulo A. Maia Neto, Stanley Coelho, Tommaso del Rosso, and Van S\'{e}rgio Alves for valuable discussions and comments.
	L.Q. and E.C.M.N. were supported by the Coordenação de Aperfeiçoamento de Pessoal de Nível Superior - Brasil (CAPES), Finance Code 001.
	D.T.A. was supported by UFPA via Licen\c{c}a Capacita\c{c}\~{a}o (Portaria 5603/2019), and thanks the hospitality of the University of Minho (Portugal), as well as that of the International Iberian Nanotechnology Laboratory (INL-Portugal).
\end{acknowledgments}
%
%%%%%%%%%%%%%%%%%%%%%%%%%%%%%%%%%%%%%%%%%%%%%%%%%%%%%%%%%%

%

\end{document}